# ViraHinter: a dual-modal artificial intelligence framework for predicting virus-host interactions

Weiqiang Bai[1,2#], Fei Wang[3#], Jialin Wang[1,2#], Sheng Xu[1,2#], Lifeng Qiao[1,4], Juan Li[5,6,7], Zhuyi Guo[8], Xiangyun Hou[3], Lei Bai[1], Bowen Zhou[1], Edward C. Holmes[9], Weifeng Shi[3,8,10*], Siqi Sun[2*]

[1] Shanghai Artificial Intelligence Laboratory, Shanghai, China

[2] Research Institute of Intelligent Complex Systems, Fudan University, Shanghai, China

[3] Ruijin Hospital, Shanghai Jiao Tong University School of Medicine, Shanghai, China

[4] School of Artificial Intelligence, Shanghai Jiao Tong University, Shanghai, China

[5] School of Clinical and Basic Medical Sciences, Shandong First Medical University & Shandong Academy of Medical Sciences, Ji'nan, China

[6] Key Laboratory of Emerging Infectious Diseases in Universities of Shandong, Shandong First Medical University & Shandong Academy of Medical Sciences, Ji'nan, China

[7] The Second Affiliated Hospital of Shandong First Medical University, Taian, China

[8] Shanghai Institute of Virology, Shanghai Jiao Tong University School of Medicine, Shanghai, China

[9] School of Medical Sciences, The University of Sydney, Sydney, NSW 2006, Australia

[10] School of Life Sciences and Biotechnology, Shanghai Jiao Tong University, Shanghai, China

[#] These authors contributed equally to this study.

[*] Corresponding author. Email: Weifeng Shi (shiwf@ioz.ac.cn), Siqi Sun (siqisun@fudan.edu.cn)

# Abstract

Protein-protein interactions (PPIs) between a virus and its host govern infection, replication, and pathogenesis. While high-throughput mapping has identified thousands of virus-host associations, much of the virus-host interactome remains uncharacterized due to the labor-intensive nature of experimental screens, the inherent difficulty in capturing transient interactions, and the limited sequence homology across divergent viral families. Here, we introduce ViraHinter, a dual-modal deep learning framework for the precise prediction of virus-host interactions and large-scale inference of interaction landscapes. ViraHinter couples a structure-generation branch with a sequence-representation branch, integrating structure-informed pair representations with ESM-derived embeddings to learn generalizable interaction rules across unseen viruses. We benchmark ViraHinter on pathogenic coronaviruses and influenza A viruses and show that it consistently outperforms RoseTTAFold2-PPI, AlphaFold 3 and RoseTTAFold2-Lite in prioritizing high-confidence candidates even under severe class imbalance and across diverse interface regimes. Notably, it successfully identifies novel functionally relevant host factors and recapitulates the structural plasticity of the complex interfaces. By intersecting predictions across multiple influenza subtypes, ViraHinter reveals 33 shared host factors, offering a roadmap for broad-spectrum antiviral discovery. ViraHinter therefore serves as a robust computational approach for studying virus-host interactions, enabling systematic screening of host factors for all known human-infecting viruses, providing new insights into the shared mechanisms of viral pathogenesis, and accelerating the discovery of novel therapeutic targets and the development of broad-spectrum antivirals.

# Introduction

Protein-protein interactions (PPIs) between a virus and its host are integral to successful viral infection, replication, and pathogenesis. By encoding a limited set of proteins that engage with a much larger host proteome, viruses hijack cellular signaling, subvert immune responses, and redirect metabolic pathways to create a permissive environment for their own propagation[1,2]. A comprehensive understanding of virus-host PPIs (vhPPIs) is therefore essential not only for elucidating the molecular mechanisms of viral infection but also for identifying novel antiviral targets[3]. In recent years, high-throughput techniques such as affinity purification-mass spectrometry (AP-MS) and proximity labeling have enabled large-scale mapping of vhPPIs, leading to the establishment of curated repositories[4,5]. Among these, IntAct offers a deeply curated collection of molecular interaction data[6]; BioGRID serves as a comprehensive repository encompassing protein, genetic, and chemical interaction datasets[7]; VirHostNet specializes in the integration and visualization of virus–host interaction networks[8]; and VirusMentha consolidates virus-host interaction data from multiple primary sources into a unified resource[9,10].

Despite their utility, existing vhPPI databases exhibit inherent limitations that constrain their translational application[11]. First, their coverage is determined by the availability of costly and labor-intensive experimental studies, leaving most of the virus-host interactome unexplored, particularly for emerging pathogens. Second, while advanced proximity-labeling systems (e.g., BioID or APEX) have begun to capture temporal snapshots of infection, their reliance on complex genetic engineering and specialized cell lines restricts their application in high-throughput comparative virology. For most emerging or non-model pathogens, the field remains constrained by static interactome maps that fail to reflect the transition between viral entry, replication, and egress[12]. There remains a critical need for a generalized computational framework that can rapidly transition from sequence to structural landscape without the prerequisite of lengthy experimental optimization. Third, because viral proteins often evolve rapidly and share limited sequence homology across families, knowledge gained from one virus, for instance, influenza A virus, does not readily transfer to another, such as coronaviruses, despite similarities in the host pathways they target. Consequently, there is a pressing need for computational approaches that can distill generalizable principles from existing interaction data and accurately predict vhPPIs across viral families in a context-aware manner.

In reality, computational methods for predicting PPIs have advanced considerably. Early deep learning approaches often rely on architectures such as long short-term memory (LSTM) networks to extract protein sequence features for PPI prediction[13,14]. However, relying solely on sequence information frequently yields poor predictive accuracy as it fails to capture the intricate spatial arrangements and physical constraints of interaction interfaces[15]. To improve performance, integrative machine learning frameworks, such as Tapioca, which employs a logistic regression-based ensemble approach, have shown promise by combining experimental dynamics data with sequence properties to map endogenous protein complexes (i.e., naturally occurring interactions within a single host species)[16]. However, these methods have not been effectively extended to virus–host interactions, partly because viral proteins frequently exploit short linear motifs or mimic host interaction interfaces, creating patterns that diverge from canonical host–host interactions and are poorly captured by homology-based methods alone[17]. Recently, structural prediction models such as AlphaFold-Multimer have demonstrated success in modeling vhPPI structures, such as those between African swine fever virus (ASFV) and human proteins[18]. However, these structural methods are computationally intensive for proteome-wide screening across divergent viral families. A model that integrates the broad relational knowledge of sequence-derived embeddings with the physical precision of structural pair representations could learn generalizable interaction rules that transcend individual virus families[19,20], enabling prediction of previously uncharacterized interactions for viruses with scarce experimental data.

Here we introduced ViraHinter, a dual-modal framework designed for high-fidelity prediction of virus-host protein-protein interactions and a systematic mapping of virus-host interaction landscapes. ViraHinter was trained on a curated, comprehensive compendium of experimentally validated interactions and integrated structure-informed pair representations with sequence-derived embeddings to learn generalizable interaction rules across viral and human proteomes. We applied the model to pathogenic coronaviruses and influenza viruses and demonstrated its robust performance across varied validation regimes. In these settings, ViraHinter successfully recovered biologically authentic interaction patterns supported by independent external evidence, confirming that the framework generalizes effectively to unseen viral families. Beyond interaction ranking, ViraHinter can uncover evolutionarily conserved binding interfaces that nominate host targets for broad-spectrum antiviral development, providing a scalable platform for characterizing the conserved host dependencies and virus-specific mechanisms that govern viral pathogenesis.

# Results

## Construction of a High-Fidelity Virus-Host Interaction Atlas

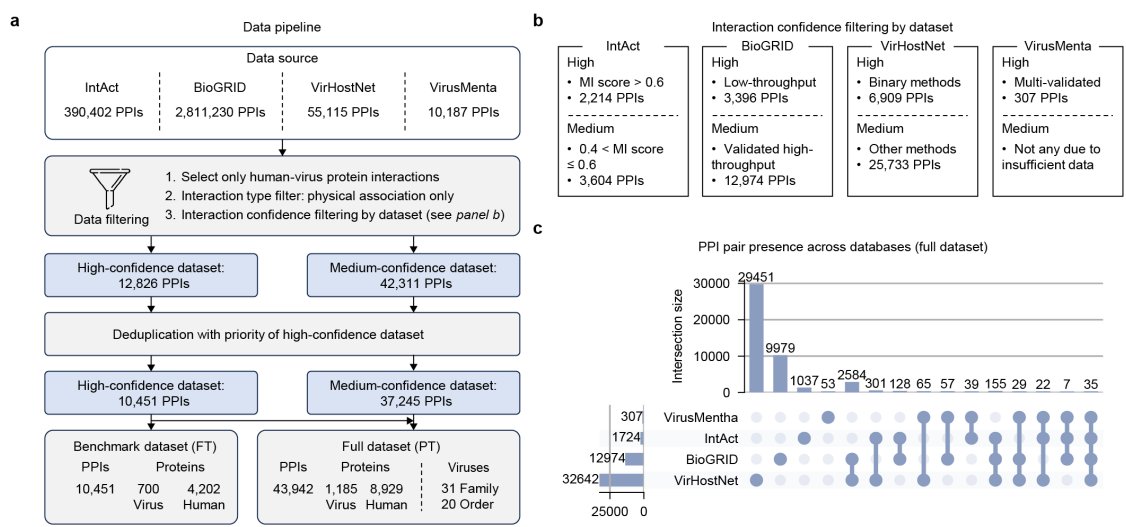

**Fig. 1 | Construction of a high-fidelity virus-host interaction atlas.**
**a**, Workflow for integrating virus-host PPIs from IntAct, BioGRID, VirHostNet, and VirusMentha. Interactions were filtered, confidence-stratified, and deduplicated to generate high-confidence and full-resource datasets for training and benchmarking. **b**, Source-specific rules used to define interaction confidence across the four databases. **c**, UpSet plot showing the overlap of virus-host PPIs across source databases.

To construct a high-fidelity atlas of virus-host protein-protein interactions, we established a staged integration workflow that harmonized the records from IntAct, BioGRID, VirHostNet and VirusMentha (**Fig. 1a**). We first retained only virus-host pairs annotated as physical associations, excluding non-physical categories such as colocalization and genetic interaction. We employed source-specific confidence strategies across repositories rather than applying a single pooled threshold (**Fig. 1b**). In IntAct, entries with an MI score > 0.6 were classified as high confidence and those with 0.4 < MI <= 0.6 as medium confidence (MI: molecular interaction confidence score summarizing evidence level; see Methods for details). In BioGRID, low-throughput physical interactions were prioritized as high confidence, whereas validated high-throughput records were retained as medium confidence. In VirHostNet, interactions supported by binary interaction methods were treated as high-confidence evidence, whereas other assay types were retained as medium confidence. For VirusMentha, because the available annotation did not support a comparable medium-confidence tier, only multi-validated records were retained as high confidence. After source-wise filtering, redundant records were collapsed with priority given to the highest-confidence annotation, yielding a curated resource that was subsequently partitioned for model development and evaluation.

After filtering and deduplication, the benchmark dataset comprised 10,451 PPIs involving 700 viral proteins and 4,202 human proteins, whereas the full dataset contained 43,942 PPIs spanning 1,185 viral proteins and 8,929 human proteins from 31 viral families across 20 viral orders (**Extended Data Table 1**). VirHostNet contributed the largest number of candidate PPIs (32,642) to the final repertoire, followed by BioGRID (12,974), IntAct (1,724), and VirusMentha (307). Most interactions were database-specific: 29,451 pairs were unique to VirHostNet, 9,979 unique to BioGRID, 1,037 unique to IntAct, and 53 unique to VirusMentha (**Fig. 1c**). The largest non-unique intersection contained 2,584 pairs shared between VirHostNet and VirusMentha, whereas all remaining pairwise or higher-order overlaps were substantially smaller. This pattern indicated that current virus–host PPI repositories were complementary rather than interchangeable, and that our unified atlas recovered the maximum number of high-confidence vhPPIs. In addition, we paired this curated positive atlas with a conservative hard-negative set generated by MMseqs2 clustering, excluding proteins from clusters containing known binders and removing candidates with sequence identity greater than 60% to any positive sample. Together, the positive atlas, hard negative set, and virus-held-out split strategy defined the data foundation for subsequent benchmarking.

**The ViraHinter Framework: Dual-Modal Fusion of Sequence and Structure**

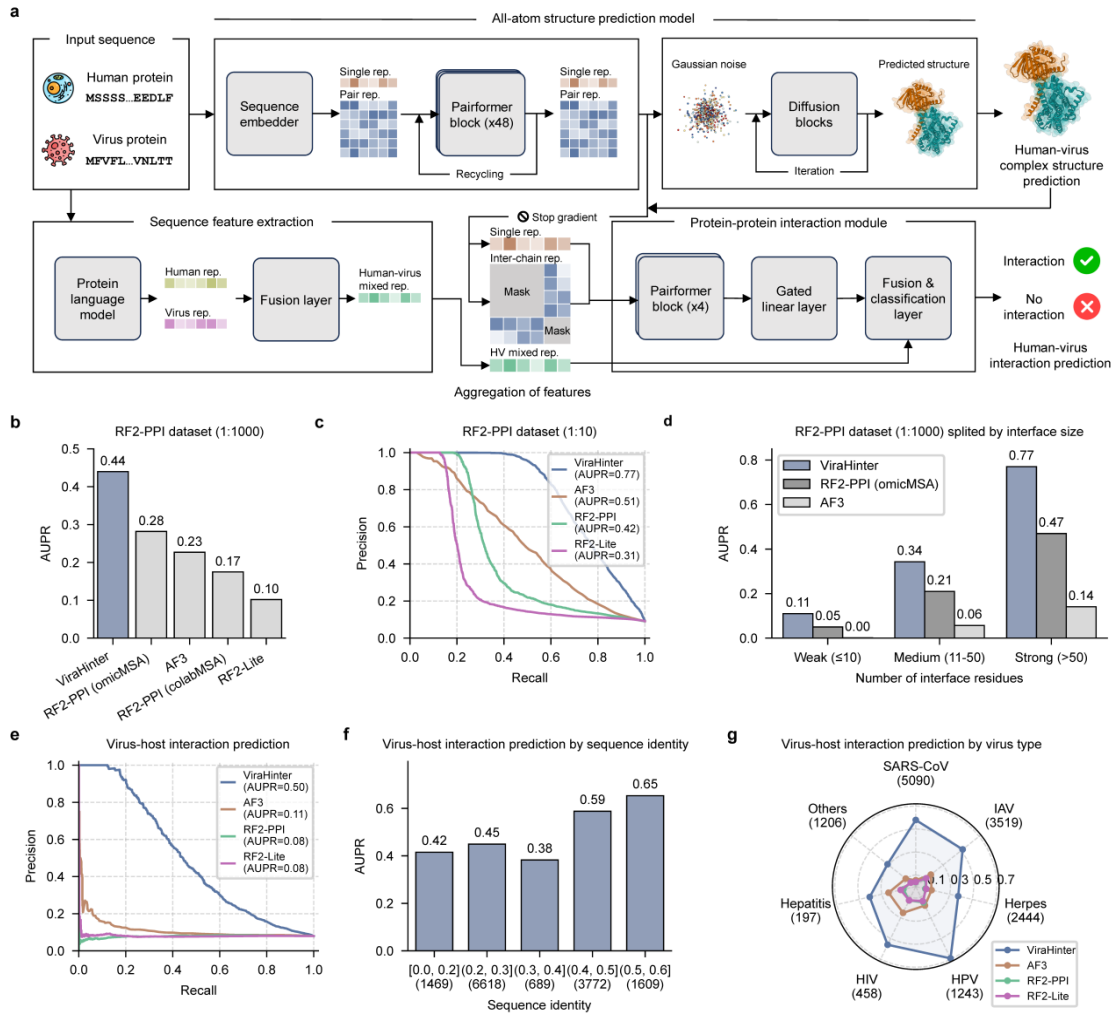

**Fig. 2 | ViraHinter architecture and benchmark performance.**
**a**, Overview of ViraHinter. A structure branch derives structure-informed single and pair representations, whereas a sequence branch extracts protein language model embeddings; these features are fused for interaction prediction. **b,c,** Performance of ViraHinter and baseline methods on RF2-PPI benchmark settings with 1:1000 and 1:10 positive-to-negative ratios. **d,** Performance on the RF2-PPI 1:1000 benchmark stratified by interface size. **e,** Performance comparison on the benchmark generated with viral-level sequence separation. Viral sequences were split so that train-test sequence identity did not exceed 60%. **f,** Performance of ViraHinter stratified by the maximum sequence identity between each test viral protein and its closest viral sequence in the training set. Test pairs were grouped into identity bins, and AUPR was calculated for each bin. **g,** Performance of ViraHinter stratified by viral group. Test pairs were grouped into SARS-CoV, IAV, Herpes, HPV, HIV, Hepatitis and Others, and AUPR was calculated for each group.

ViraHinter is a dual-modal architecture that couples a structure-generation branch with a sequence-representation branch and routes the fused features to both structure and interaction heads (**Fig. 2a**). Given paired human and viral sequences, the structural branch first uses a sequence embedder and a 48-block Pairformer stack to produce single and pair representations, which are then refined by an iterative diffusion module to generate all-atom virus-host complex structures. In parallel, the sequence branch extracts ESM-derived embeddings for the human and viral partners and combines them through a fusion layer to form a mixed virus-host representation. To focus the interaction predictor on intermolecular compatibility rather than monomeric folding signals, the pair representation is masked to suppress intra-chain information before feature aggregation. Structure-derived features are then transferred to the interaction branch through a stop-gradient connection and jointly processed with the sequence-derived representation by four additional Pairformer blocks, a gated linear layer, and a final fusion-classification layer, which outputs the interaction probability.

This design enables ViraHinter to leverage explicit geometric context when reliable pairwise structural cues are available, while preserving sequence-derived information for cases in which structural evidence is weak, such as rapidly evolving or partially disordered viral proteins. The model was trained for virus–host interaction prediction using a structure-aware backbone. Specifically, the structure module was initialized from IntFold[21,22], after which the model was [22]subset. By integrating both modalities within a single interaction head, ViraHinter ranks candidate virus–host pairs while simultaneously generating the corresponding complex structures, rather than relying on a post hoc combination of separate predictors.

**Benchmarking and Performance Evaluation**

We next evaluated ViraHinter under the benchmarking settings introduced in the original RF2-PPI study[23], which tested performance under severe class imbalance and across different interface-size regimes (**Fig. 2b–d**). Under the stringent 1:1,000 positive-to-negative ratio intended to approximate the search space of large-scale screening, ViraHinter achieved an AUPR of 0.44, outperforming RF2-PPI[23] (omicMSA, 0.28), AlphaFold 3(AF3)[24] (0.23), RF2-PPI (colabMSA, 0.17)[23] and RF2-Lite[25] (0.10) (**Fig. 2b**). This advantage was maintained in the less imbalanced 1:10 setting, in which ViraHinter reached an AUPR of 0.77, compared with 0.51 for AF3, 0.42 for RF2-PPI and 0.31 for RF2-Lite (**Fig. 2c**). To determine whether the improvement was restricted to large and readily resolved complexes, we further stratified the 1:1,000 benchmark by interface size, following the evaluation protocol of the RF2-PPI[23] study. ViraHinter ViraHinter consistently outperformed baseline methods across interfaces of different sizes, achieving AUPRs of 0.11, 0.34, and 0.77 on weak (≤10 residues), medium (11–50 residues), and strong (>50 residues) interfaces, respectively. By comparison, RF2-PPI (omicMSA) achieved 0.05, 0.21, and 0.47, and AF3 achieved 0.00, 0.06, and 0.14 across the same bins (**Fig. 2d**). Interface size was defined following the RF2-PPI protocol, based on the number of residues at the binding site of the experimentally determined protein–protein complex. Together, these results showed that ViraHinter consistently outperformed existing methods across the benchmark regimes established by RF2-PPI, including settings with extreme class imbalance and limited interface size.

We next assessed ViraHinter on additional virus-host benchmarks constructed with data-partitioning schemes designed to reduce viral leakage **(Fig. 2e-g)**. In the benchmark partitioned by viral sequence identity, where test viral proteins shared at most 60% sequence identity with the closest viral proteins in the training set, ViraHinter yielded an AUPR of 0.50, a 4.5-fold improvement over AF3 (0.11) and more than 6-fold over both RF2-PPI and RF2-Lite (0.08) **(Fig. 2e)**. This margin was notably wider than that observed in the RF2-PPI benchmark settings **(Fig. 2b-d)**, suggesting that the dual-modal architecture is particularly effective when sequence homology to training data is limited, precisely the scenario encountered with emerging or understudied pathogens. Because comparable training-set identities could not be defined uniformly for all baseline methods, we next examined ViraHinter alone as a function of viral sequence identity to the training set **(Fig. 2f)**. Across all populated identity bins, ViraHinter retained measurable predictive power, with AUPRs of 0.42 for test viral proteins with 10–20% identity to the closest training viral sequence, 0.45 for 20–30%, 0.38 for 30–40%, 0.59 for 40–50% and 0.65 for 50–60%.

Test pairs were grouped into viral categories, including SARS-CoV (including SARS-CoV-1 and SARS-CoV-2), Influenza A Virus (IAV), Human Immunodeficiency Virus (HIV), Human Papillomavirus (HPV), herpesviruses(Herpes), hepatitis viruses(Hepatitis) and an Other category **(Extended Data Table 2)**, and ViraHinter consistently outperformed AF3, RF2-PPI and RF2-Lite across all groups in this benchmark **(Fig. 2g)**. ViraHinter achieved the highest AUPR on HPV (0.69), followed by SARS-CoV (0.58), HIV (0.56) and IAV (0.52), whereas performance was lower for the hepatitis virus group (0.41), the herpesvirus group (0.38) and the Other category (0.31), but still remained clearly above the baselines .These analyses, together with the RF2-PPI benchmarks above, indicated that ViraHinter generalized across diverse viral groups and retained predictive utility even for divergent viral sequences with low training-set homology.

**Validation of ViraHinter against experimental host-coronavirus interactomes**

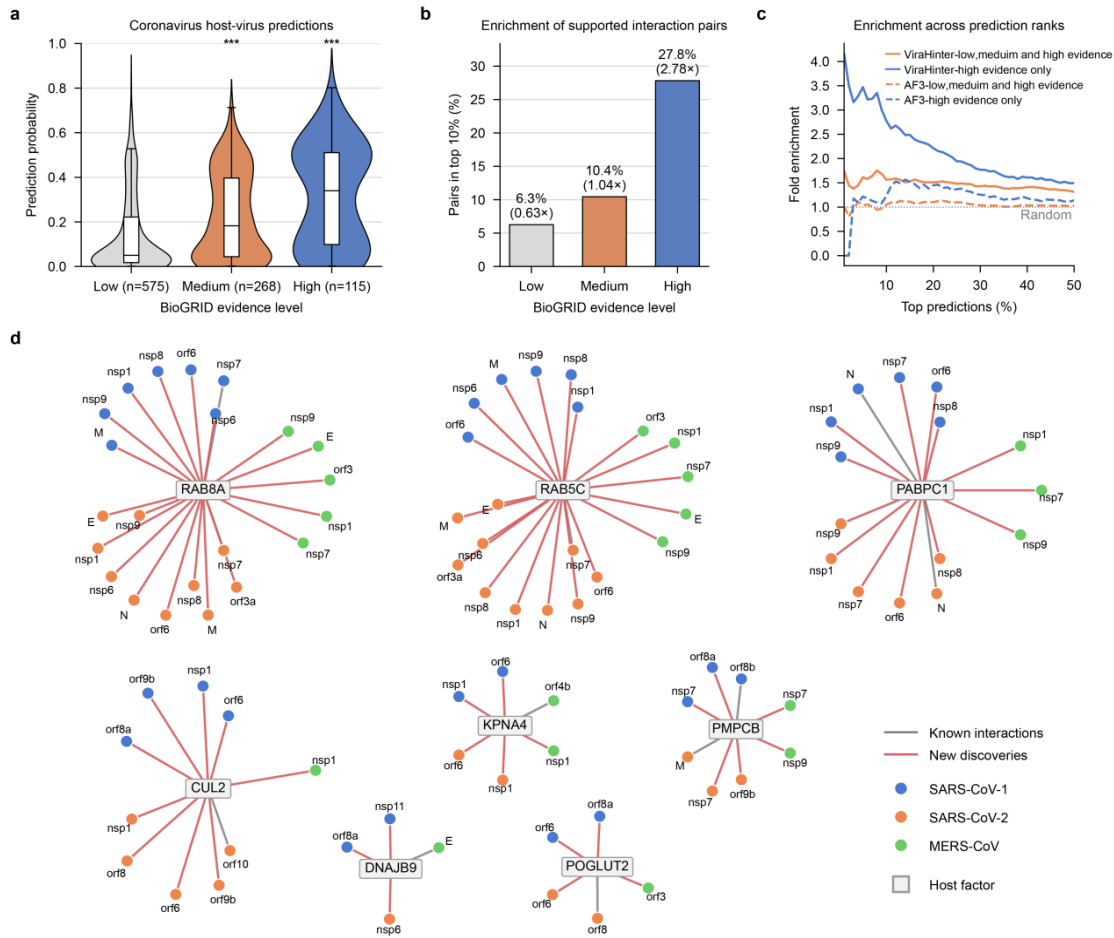

**Fig. 3 | Evidence-aware validation on coronavirus interactions.**

**a,** Distribution of ViraHinter prediction probabilities for 958 candidate coronavirus–host interaction pairs, stratified by BioGRID confidence levels: Low (no qualifying evidence, n=575), Medium (high-throughput evidence, n=268), and High (experimentally verified, n=115). Box plots represent the median and interquartile range (IQR); whiskers extend to 1.5 times×IQR. *** $P < 0.001$, Kruskal-Wallis test. **b,** Fraction and enrichment of the supported interaction pairs within the top 10% of ViraHinter predictions. Percentages indicate the proportion of known pairs recovered from the total available pairs in each category; values in parentheses represent fold enrichment over random expectation. **c,** Fold enrichment of the supported pairs across a continuous range of prediction-rank thresholds (top 1% to 50%). ViraHinter (solid lines) consistently maintains a superior enrichment profile over AF3 (dashed lines) for both high-confidence (blue) and low and no-confidence (orange) interactions. Dotted gray line represents random expectation (fold enrichment =1). **d,** Predicted shared interaction networks for eight prioritized host factors (RAB8A, RAB5C, PABPC1, DNAJB9, KPNA4, POGLUT2, CUL2 and PMPCB) across the three pathogenic coronaviruses. Viral bait proteins are colored by virus: SARS-CoV-1 (blue), SARS-CoV-2 (orange), and MERS-CoV (green). Gray edges represent known interactions documented in BioGRID. Red edges denote *de novo* discoveries predicted by ViraHinter that expand the binding profile of these factors across the three coronaviruses.

To evaluate the predictive fidelity of ViraHinter, we benchmarked the framework against 958 candidate virus–host protein pairs from SARS-CoV-2, SARS-CoV-1, and MERS-CoV, originally identified via high-throughput AP-MS[5]. After removing training set overlaps to ensure a de-leaked interaction-level validation, we categorized the benchmarked pairs using external BioGRID records: (i) high-throughput (HT) evidence, (ii) low-throughput (LT) evidence, and (iii) no qualifying evidence **(Extended Data Table 3)**. Predicted interaction probabilities differed significantly across these categories, with the highest scores assigned to the pairs with experimentally verified (LT) evidence and the lowest scores assigned to the pairs without available evidence **(Fig. 3a)**. This ordering indicated that ViraHinter preferentially ranked coronavirus interaction candidates with stronger independent support toward the top of the predicted probability list. The practical utility of ViraHinter lay in its ability to prioritize candidates for experimental validation. In an enrichment analysis of the top 10% of the predictions **(Extended Data Table 3)**, we observed a substantial overrepresentation of the validated interactors. Specifically, ViraHinter recovered 27.8% of experimentally verified interactions, representing a 2.78-fold enrichment over random expectation **(Fig. 3b)**. Notably, pairs lacking evidence were markedly depleted (0.63-fold enrichment), whereas pairs with HT evidence showed only marginal gains **(1.04-fold; Fig. 3b)**. This sensitivity to "evidence quality" rather than simple record presence allows ViraHinter to effectively identify high-confidence targets. Robustness analysis revealed that the supported interactions remain enriched across the entire spectrum of ranking cutoffs **(Fig. 3c).** The enrichment was most pronounced for experimentally verified pairs, peaking at ~4.0-fold within the highest-confidence window (top 1%-5%) **(Fig. 3c).** When benchmarked against AF3, ViraHinter consistently maintained a superior enrichment profile, particularly in the critical high-confidence region **(Fig. 3c).** For both "high-confidence" (LT evidence) and the broader "low and no evidence"(HT evidence and no qualifying evidence) categories, ViraHinter's recovery outperformed AF3's across the majority of the ranking window **(Fig. 3c).** These results underscored the robust, evidence-aware ranking capability of ViraHinter that prioritizes reliable, physically verified PPIs for downstream study.

We next utilized ViraHinter to uncover conserved host dependencies potentially missed by traditional mapping. By prioritizing the top 100 host genes **(Extended Data Table 4)**, the framework identified eight with predicted binding profiles that diverged significantly from existing records **(Fig. 3d)**. Most notably, ViraHinter found seven factors (PABPC1, RAB8A, RAB5C, CUL2, KPNA4, PMPCB, and POGLUT2) to be potential pan-coronavirus targets as they were shared by all the three coronaviruses **(Fig. 3d)**. Network analysis revealed extensive "New discoveries" (red edges) for these factors, such as RAB8A and RAB5C, which were predicted to engage a broad array of viral baits as universal machinery for trafficking or replication **(Fig. 3d)**. Additionally, DNAJB9 exhibited an expanded profile, transitioning from a single-virus to a pan-coronavirus interactor **(Fig. 3d)**. These high-priority candidates represented functional interactors whose broader roles were likely masked in physical repositories due to transient binding.To provide a mechanistic basis for these shared dependencies, we modelled the structural interfaces between RAB8A and NSP7 from all three coronaviruses. NSP7 is a critical component of the viral replication-transcription complex (RTC) **(Extended Data Fig. 2a).** Structural analysis revealed a strikingly conserved binding mode across these divergent coronaviruses. Despite significant sequence variations between the NSP7 orthologs, the predicted binding interfaces were virtually coincident, converging upon an identical functional cleft on RAB8A **(Extended Data Fig. 2a)**. The nearly identical spatial orientation and high geometric complementarity suggested that the NSP7-RAB8A interaction axis was evolutionarily conserved, implying that RAB8A is an indispensable cellular node and hence likely a robust, broad-spectrum therapeutic target resistant to mutational escape.

**ViraHinter identifies conserved and subtype-specific host-influenza interactomes**

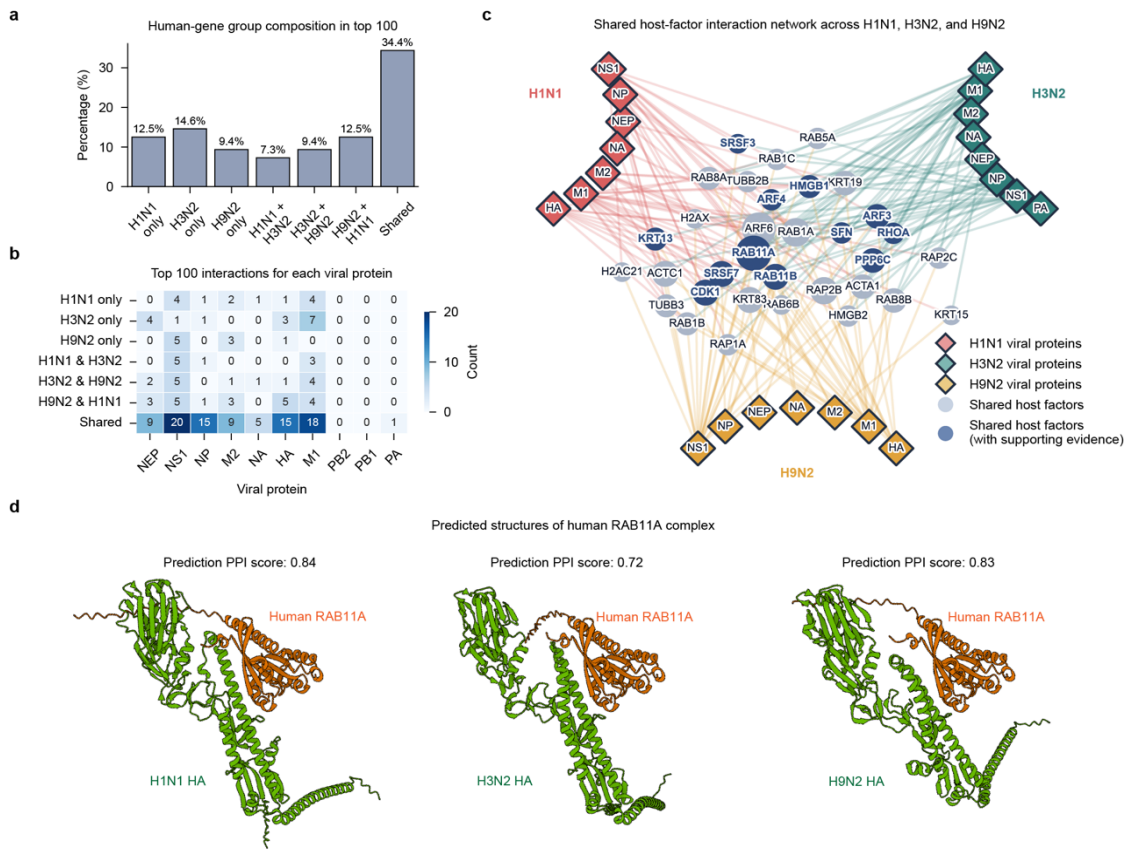

**Fig. 4 | Identification and network characterization of shared host factors across H1N1, H3N2, and H9N2 influenza viruses.**
ViraHinter was applied to H1N1, H3N2 and H9N2 in a virus-held-out setting; the evaluated influenza viruses were excluded from training, and highly similar external records were also removed during evidence lookup. **a,** Gene-level composition of the predicted host factors in the top 100 ranked candidates. Categories represent subtype-specific host factors (H1N1, H3N2, or H9N2 only), pairwise-shared host factors, and factors shared by all the three subtypes (Shared). Percentages are calculated relative to all host genes. **b,** Heatmap showing the distribution of host genes across individual viral baits in the top 100 predictions, stratified by the categories defined in panel b. Row categories indicate the specificity of the host factor (e.g., Shared vs. H3N2 only). **c,** Shared host-factor interaction network formed by the core host genes targeted by all three influenza subtypes in the top 100 predictions. Colored diamond nodes denote viral proteins from H1N1 (red), H3N2 (green), and H9N2 (yellow); light blue circles represent shared human proteins, while dark blue circles indicate shared host factors with independent supporting evidence (e.g., SFN, RAB11A, CDK1). Edge thickness or length reflects the ViraHinter interaction score. **d,** Structural conservation of the HA-RAB11A interaction interface. Predicted complex structures of human RAB11A (orange) in association with HA proteins (green) from H1N1, H3N2, and H9N2 strains are shown.

Influenza A viruses are characterized by rapid mutation and antigenic shifts that create a diverse array of antigenically distinct subtypes[26]. To investigate whether human and avian-origin influenza viruses exploit a shared set of host factors, we performed a comparative analysis using three representative subtypes: two human-infecting subtypes, the seasonal A/Nagasaki/07N020/2008 (H1N1) strain and the seasonal A/China/JN22022/2022 (H3N2) strain, alongside an avian-origin A/chicken/Korea/KNUGJ09/2009 (H9N2) strain. Utilizing ViraHinter, we conducted an exhaustive screen for the 10 essential proteins (PB2, NS1, NEP, M2, M1, NP, HA, PA, PB1, and NA) from the isolated H3N2 strain (isolated and sequenced in our laboratory in 2025), along with H1N1 and H9N2 against the human proteome utilizing ViraHinter. To ensure rigorous validation, the viruses evaluated were excluded from training at the subtype level, and external evidence was de-leaked by removing records that overlapped with or shared high sequence identity with the training set. Comparative analysis revealed a stable pan-influenza core within the host-targeted interaction landscape **(Fig. 4a and Extended Data Fig. 3a)**. In the Top 100 predictions, 34.4% of the host factors were shared across all three subtypes **(Fig. 4a)**, which expanded to 56.5% in the top 1000, highlighting the framework's consistence. Concurrently, distinct strain-specific profiles were identified: in the top 100, H1N1-only interactions accounted for 12.5%, compared to 14.6% for H3N2 and 9.4% for H9N2 **(Fig. 4a)**. As the evaluation depth increased to the Top 1000, these lineage-specific ratios shifted to 5.3% for H1N1, 5.1% for H3N2, and 4.8% for H9N2 **(Extended Data Fig. 3a)**. The observation that shared interactions consistently dwarf strain-specific counts highlights ViraHinter's ability to identify the "evolutionary bottleneck" of host factors essential for pan-influenza pathogenesis, effectively filtering out stochastic noise to reveal a robust core of conserved dependencies.

To distinguish high-affinity primary targets from broader functional networks, we performed a systematic deconvolution of the predicted interactomes **(Fig. 4b)**. In the Top 100, the shared interactome was anchored by "hub" proteins with consistent targeting across subtypes: NS1 emerged as the primary anchor (n=20 common factors), followed by M1 (n=18), NP (n=15), and HA (n=15) **(Fig. 4b)**. At this high-confidence threshold, the polymerase subunits (**PB2** and **PB1**) exhibited minimal shared interactions **(Fig. 4b)**. This observation is consistent with their roles as highly conserved replicase proteins, suggesting that their primary high-affinity targets are restricted to highly specialized cellular nodes or involve transient associations required for the precision of the viral replication cycle. Interestingly, M1 exhibited a lineage-specific signature in H3N2 strain (n=7), potentially reflecting subtype-specific variations in how the matrix protein engages cellular machinery **(Fig. 4b)**. Expanding the analysis to the top 1000 interactions revealed a dramatic scaling of the shared core **(Extended Data Fig. 3b)**. The shared interactors for NS1 increased tenfold (n=196), with significant expansions for HA (n=173), NEP (n=167) and M1 (n=148). Crucially, the polymerase subunits (PB1, PA, and PB2) emerged as robust shared hubs at this depth. While the requirement of a conserved cellular network for viral replication is well-established, ViraHinter's sensitivity at this threshold reveals the full scale of this shared infrastructure, capturing the broader array of transient or lower-affinity associations that are often underrepresented in high-stringency experimental datasets. The predominance of the shared counts over subtype-specific ones (e.g., NS1 shared 196 vs. H3N2-only 1) underscores ViraHinter's ability to identify evolutionarily conserved characterizations in IAV pathogenesis.

By intersecting the top 100 interactors for H1N1, H9N2, and H3N2 **(Extended Data Table 5)**, we identified a coherent network of 33 common host factors targeted by all three strains **(Fig. 4c, Extended Data Table 5)**. Functional mapping **(Extended Data Fig. 4d)** revealed that these factors are distributed across five primary modules: (i) vesicular transport, (ii) cytoskeleton, (iii) chromatin and DNA damage, (iv) RNA metabolism and splicing, and (v) signal transduction and cycle regulation. Several key nodes exhibit critical cross-functional roles; notably, CDK1 bridges signal transduction with chromatin integrity[27,28], while HMGB1 and HMGB2 coordinate RNA transcription and splicing with chromatin-mediated DNA repair[29,30]. To validate the biological relevance of this predicted core, we cross-referenced these candidates with independent experimental literature. Among the 33 core factors, 12 host factors (including SFN, RAB11A, RAB11B, ARF3, CDK1, SRSF3, SRSF7, RHOA, HMGB1, ARF4, PPP6C and KRT13) currently lack physical interaction evidence with IAV-encoded proteins in the virus-host interaction databases but have been documented to regulate host immune responses specifically during IAV infection[31-42] **(Extended Data Table 6)**. A further 18 factors (KAT15, ARF6, RAP2C, RAP2B, RAB1B, RAB1A, RAB5A, KRT19, ACTA1, HMGB2, RAB8B, ACTC1, H2AX, RAP1A, RAB8A, TUBB3, TUBB2B and RAB6B) have been reported to modulate immune responses to other viruses[33][43-59] **(Extended Data Table 6)**. This finding demonstrates that ViraHinter's dual-modal architecture effectively compensates for experimental limitations by identifying elusive but functionally essential host targets through structural and sequence-derived signatures.

Because the influenza virus analysis involved proteome-scale screening against the full human protein repertoire, we adopted a multi-stage candidate selection strategy to make large-scale evaluation computationally tractable. Candidate virus-host pairs were first prescreened using a lightweight sequence-only model (see Methods for details) and were then rescored with the full ViraHinter model. For downstream prioritization, the top 100 candidates were further reranked using a composite score integrating the full-model interaction score, ipTM and the sequence-only score, thereby balancing predicted interaction strength, structural confidence and sequence-based support. This formula integrates the model's structural complementarity features (see Methods for details), the interface confidence (ipTM), and evolutionary sequence conservation into a single prioritized metric **(Extended Data Table 5)**.

Finally, atomic-level modeling of the interaction between RAB11A and HA from H3N2, H1N1, and H9N2 provided a mechanistic rationale for these shared dependencies **(Fig. 4d)**. RAB11A emerged as a cornerstone of the shared network. Despite the extensive sequence plasticity and antigenic diversity among HA subtypes, ViraHinter predicted a strikingly conserved binding mode, with the HA proteins from all three subtypes converging on an identical functional cleft on the RAB11A surface **(Fig. 4d and Extended Data Fig. 3c)**. This structural consistency suggested that RAB11A may represent an essential evolutionary bottleneck for influenza viruses. In particular, the requirement to maintain high-affinity docking with the RAB11A machinery imposes a strong negative selection pressure on the interaction interface.

## Discussion

We present ViraHinter, a dual-modal framework that integrates structure-informed pair representations with sequence-derived features for virus-host protein-protein interaction prediction. By combining a curated multi-source interaction atlas with staged training across general protein complexes, broad PPI data, and progressively refined virus-host subsets, ViraHinter supports both candidate ranking and structure-aware interpretation within a unified model. The benchmarking analyses showed that this architecture not only outperformed state-of-the-art models like AF3[24] and RF2-PPI in large-scale ranking scenarios, particularly under severe class imbalance, but also provided a mechanistic lens through which to interpret conserved pathogenic strategies.

A primary contribution of ViraHinter was its ability to "rescue" functionally essential interactions that were generally missed by traditional proteomics. Physical interaction repositories, while invaluable, were often biased toward stable complexes, failing to capture the transient or low-affinity associations that define viral hijacking. In our coronavirus analysis, ViraHinter prioritized a core set of shared host factors, including RAB8A, RAB5C, and CUL2; although many lacked prior physical evidence, the model identified them as pan-viral anchors. Similarly, in the influenza analysis, the framework identified 33 common host factors, 12 of which (including SFN and RAB11A) have been shown to regulate host immunity during IAV infection despite the absence of physical interaction records in existing vhPPI databases[31,37]. This "functional rescue" underscored the utility of ViraHinter's dual-modal design, which leveraged structural compatibility to identify elusive yet biologically authentic targets.

Of particular note was that ViraHinter highlighted the existence of the evolutionary conservation of conserved aspects of interaction between influenza virus and the host proteome. While surface proteins like HA and NA undergo rapid antigenic evolution, internal replication proteins, and the host response to them, remained highly conserved. ViraHinter captured this by identifying a stable pan-influenza core anchored by conserved proteins such as M1 and NS1. Crucially, comparative structural modeling of two key interaction axes, RAB11A in complex with HA from three influenza subtypes (H1N1, H3N2, and H9N2) and RAB8A with NSP7 from three pathogenic coronaviruses (SARS-CoV-1, SARS-CoV-2, and MERS-CoV), revealed strikingly coincident binding modes. Despite the evolutionary divergence of these viral proteins, they converge on virtually identical functional clefts on their respective host targets,

RAB11A and RAB8A. The ability of ViraHinter to capture these structurally "locked" interfaces, even in the absence of high sequence homology, underscores its utility in identifying robust, broad-spectrum therapeutic targets.. Such insights are instrumental for developing interventions that are inherently resistant to the rapid mutational escape typical of seasonal and pandemic influenza strains. Furthermore, ViraHinter also demonstrated robust generalization in data-scarce scenarios, manifest in its ability to characterize the H3N2 strain identified in our laboratory and the data-limited H1N1 and H9N2 viruses, such that it is ideally suited to monitoring emerging zoonotic threats.

Although ViraHinter exceled at interaction ranking and binding region identification, residue-level precision in side-chain packing, it clearly requires further refinement to match high-resolution cryo-EM data. Future iterations will incorporate iterative wet-lab validation to fill existing data gaps and hence reduce bias in the structural data sets. By bridging structural biology with functional proteomics, ViraHinter represents a powerful platform for comparative virology. enabling the rapid identification of pathogenic mechanisms and therapeutic targets across different human-infecting viral families.

# References

1. Kruse, T.*, et al.* Large scale discovery of coronavirus-host factor protein interaction motifs reveals SARS-CoV-2 specific mechanisms and vulnerabilities. *Nat Commun* **12**, 6761 (2021).
2. Stukalov, A.*, et al.* Multilevel proteomics reveals host perturbations by SARS-CoV-2 and SARS-CoV. *Nature* **594**, 246-252 (2021).
3. Gordon, D.E.*, et al.* A SARS-CoV-2 protein interaction map reveals targets for drug repurposing. *Nature* **583**, 459-468 (2020).
4. Zhao, X., Zheng, X., Liang, Z., Zheng, C. & Shang, P. Identification of Virus-Host Protein Interactions Via Proteomic Techniques. *Methods Mol Biol* **2940**, 151-163 (2025).
5. Gordon, D.E.*, et al.* Comparative host-coronavirus protein interaction networks reveal pan-viral disease mechanisms. *Science* **370**(2020).
6. Del Toro, N.*, et al.* The IntAct database: efficient access to fine-grained molecular interaction data. *Nucleic Acids Res* **50**, D648-D653 (2022).
7. Chatr-Aryamontri, A.*, et al.* The BioGRID interaction database: 2015 update. *Nucleic Acids Res* **43**, D470-478 (2015).
8. Navratil, V.*, et al.* VirHostNet: a knowledge base for the management and the analysis of proteome-wide virus-host interaction networks. *Nucleic Acids Res* **37**, D661-668 (2009).

9.  Saha, D., Iannuccelli, M., Brun, C., Zanzoni, A. & Licata, L. The Intricacy of the Viral-Human Protein Interaction Networks: Resources, Data, and Analyses. *Front Microbiol* **13**, 849781 (2022).
10. Calderone, A., Licata, L. & Cesareni, G. VirusMentha: a new resource for virus-host protein interactions. *Nucleic Acids Res* **43**, D588-592 (2015).
11. Iuchi, H.*, et al.* Bioinformatics approaches for unveiling virus-host interactions. *Comput Struct Biotechnol J* **21**, 1774-1784 (2023).
12. Hashimoto, Y., Sheng, X., Murray-Nerger, L.A. & Cristea, I.M. Temporal dynamics of protein complex formation and dissociation during human cytomegalovirus infection. *Nat Commun* **11**, 806 (2020).
13. Noor, F. & Tahir Ul Qamar, M. Comprehensive review and assessment of machine learning approaches for host-pathogen protein-protein interaction prediction. *Brief Bioinform* **27**(2026).
14. Soleymani, F., Paquet, E., Viktor, H., Michalowski, W. & Spinello, D. Protein-protein interaction prediction with deep learning: A comprehensive review. *Comput Struct Biotechnol J* **20**, 5316-5341 (2022).
15. Mou, M., Zhang, Z., Pan, Z. & Zhu, F. Deep Learning for Predicting Biomolecular Binding Sites of Proteins. *Research (Wash D C)* **8**, 0615 (2025).
16. Reed, T.J., Tyl, M.D., Tadych, A., Troyanskaya, O.G. & Cristea, I.M. Tapioca: a platform for predicting de novo protein-protein interactions in dynamic contexts. *Nat Methods* **21**, 488-500 (2024).
17. Bogdanow, B.*, et al.* Structural host-virus interactome profiling of intact infected cells. *Nat Commun* **16**, 6713 (2025).
18. Fenster, J.A.*, et al.* African Swine Fever Virus Protein-Protein Interaction Prediction. *Viruses* **16**(2024).
19. Wang, X.*, et al.* DeepHVI: A multimodal deep learning framework for predicting human-virus protein-protein interactions using protein language models. *Biosaf Health* **7**, 257-266 (2025).
20. Lin, Z.*, et al.* Evolutionary-scale prediction of atomic-level protein structure with a language model. *Science* **379**, 1123-1130 (2023).
21. Qiao, L., et al. IntFold: A Controllable Foundation Model for General and Specialized Biomolecular Structure Prediction. Preprint at https://arxiv.org/abs/2507.02025 (2025).
22. Qiao, L.f., et al. IntelliFold-2: Surpassing AlphaFold 3 via Architectural Refinement and Structural Consistency. Preprint at bioRxiv https://doi.org/10.64898/2026.02.09.704787 (2026).
23. Zhang, J.*, et al.* Predicting protein-protein interactions in the human proteome. *Science* **390**, eadt1630 (2025).
24. Abramson, J.*, et al.* Accurate structure prediction of biomolecular interactions with AlphaFold 3. *Nature* **630**, 493-500 (2024).
25. Humphreys, I.R.*, et al.* Protein interactions in human pathogens revealed through deep learning. *Nat Microbiol* **9**, 2642-2652 (2024).
26. Petrova, V.N. & Russell, C.A. The evolution of seasonal influenza viruses. *Nat Rev Microbiol* **16**, 60 (2018).
27. Fernandez-Casanas, M.*, et al.* Centromere protection requires strict mitotic inactivation of the Bloom syndrome helicase complex. *Nat Commun* **16**, 7832 (2025).

28. Khan, A., *et al.* A SETD2-CDK1-lamin axis maintains nuclear morphology and genome stability. *Nat Cell Biol* **27**, 1327-1341 (2025).
29. Mandke, P. & Vasquez, K.M. Interactions of high mobility group box protein 1 (HMGB1) with nucleic acids: Implications in DNA repair and immune responses. *DNA Repair (Amst)* **83**, 102701 (2019).
30. Reeves, R. High mobility group (HMG) proteins: Modulators of chromatin structure and DNA repair in mammalian cells. *DNA Repair (Amst)* **36**, 122-136 (2015).
31. Ganti, K., Han, J., Manicassamy, B. & Lowen, A.C. Rab11a mediates cell-cell spread and reassortment of influenza A virus genomes via tunneling nanotubes. *PLoS Pathog* **17**, e1009321 (2021).
32. Guo, J., *et al.* A genome-wide base-editing screen uncovers a pivotal role of paxillin delta ubiquitination in influenza virus infection. *Cell Rep* **44**, 115748 (2025).
33. Huang, Y., Urban, C., Hubel, P., Stukalov, A. & Pichlmair, A. Protein turnover regulation is critical for influenza A virus infection. *Cell Syst* **15**, 911-929 e918 (2024).
34. Kazer, S.W., *et al.* Primary nasal influenza infection rewires tissue-scale memory response dynamics. *Immunity* **57**, 1955-1974 e1958 (2024).
35. Li, M.Y., *et al.* ARF4-mediated intracellular transport as a broad-spectrum antiviral target. *Nat Microbiol* **10**, 710-723 (2025).
36. Sun, H., *et al.* Genome-wide CRISPR screen identifies STK11 as a critical regulator of sialic acid clusters important for influenza A virus attachment. *J Adv Res* (2025).
37. Waqas, F.H., *et al.* NRF2 activators inhibit influenza A virus replication by interfering with nucleo-cytoplasmic export of viral RNPs in an NRF2-independent manner. *PLoS Pathog* **19**, e1011506 (2023).
38. Zhang, Z., Lu, B. & Zeng, B. ARF3 knockdown inhibits influenza a virus and virus-induced pneumonia. *Virus Genes* **61**, 554-561 (2025).
39. Zhao, L., *et al.* Host-specific SRSF7 regulates polymerase activity and replication of influenza A virus. *Microbes Infect* **26**, 105401 (2024).
40. Zhao, L., *et al.* SRSF3 facilitates replication of influenza A virus via binding and promoting the transport of viral mRNA. *Vet Microbiol* **266**, 109343 (2022).
41. Zhao, L., *et al.* The CDK1 inhibitor, Ro-3306, is a potential antiviral candidate against influenza virus infection. *Antiviral Res* **201**, 105296 (2022).
42. Turner, A.H., et al. Rab11B is required for binding and entry of recent H3N2, but not H1N1, influenza A isolates. Preprint at bioRxiv https://doi.org/10.1101/2025.04.23.650275 (2025).
43. Abbasi, A., *et al.* Serum extracellular vesicle RNA profiles in long COVID: insights from exercise-induced gene modulation. *Sci Rep* **16**, 3469 (2026).
44. Davidson, L.A., *et al.* Mining bulk transcriptomic datasets identifies inflammasome activation and antigen presentation as key novel mechanisms of BK polyomavirus-associated nephropathy. *J Pathol* **268**, 288-297 (2026).
45. Jiao, H., *et al.* Deep Insight Into Long Non-coding RNA and mRNA Transcriptome Profiling in HepG2 Cells Expressing Genotype IV Swine Hepatitis E Virus ORF3. *Front Vet Sci* **8**, 625609 (2021).

46. Kang, H., Kang, T., Jackson, L., Murphy, A. & Nitta, T. Evidence for Involvement of ADP-Ribosylation Factor 6 in Intracellular Trafficking and Release of Murine Leukemia Virus Gag. *Cells* **13**(2024).
47. Li, C.*, et al.* RAB1A promotes hepatitis B virus replication by enhancing PPARalpha-mediated viral transcription and inducing ULK1-mediated autophagy. *Int J Biol Macromol* **321**, 146301 (2025).
48. Li, M.*, et al.* Serological and Molecular Characterization of Hepatitis B Virus Infection in Gastric Cancer. *Front Cell Infect Microbiol* **12**, 894836 (2022).
49. Li, W.*, et al.* Characteristic of HPV Integration in the Genome and Transcriptome of Cervical Cancer Tissues. *Biomed Res Int* **2018**, 6242173 (2018).
50. Li, W.*, et al.* Oncogenic KSHV-encoded interferon regulatory factor upregulates HMGB2 and CMPK1 expression to promote cell invasion by disrupting a complex lncRNA-OIP5-AS1/miR-218-5p network. *PLoS Pathog* **15**, e1007578 (2019).
51. Li, X.*, et al.* Proteomic characteristics of the treatment trajectory of patients with COVID-19. *Arch Virol* **169**, 84 (2024).
52. Ming, X.*, et al.* Pseudorabies virus kinase UL13 phosphorylates H2AX to foster viral replication. *FASEB J* **36**, e22221 (2022).
53. Salyers, Z.R., Coleman, M., Le, D. & Ryan, T.E. AAV-mediated expression of PFKFB3 in myofibers, but not endothelial cells, improves ischemic muscle function in mice with critical limb ischemia. *Am J Physiol Heart Circ Physiol* **323**, H424-H436 (2022).
54. Xin, J.*, et al.* Exploring the antiviral potential of shikimic acid against Chikungunya virus through network pharmacology, molecular docking, and in vitro experiments. *Front Vet Sci* **12**, 1524812 (2025).
55. Yang, W.*, et al.* SNX32 is a host restriction factor that degrades African swine fever virus CP204L via the RAB1B-dependent autophagy pathway. *J Virol* **98**, e0159923 (2024).
56. Ying, G.*, et al.* Small GTPases Rab8a and Rab11a Are Dispensable for Rhodopsin Transport in Mouse Photoreceptors. *PLoS One* **11**, e0161236 (2016).
57. Yoo, S.Y., Jeong, S.N., Kang, J.I. & Lee, S.W. Chimeric Adeno-Associated Virus-Mediated Cardiovascular Reprogramming for Ischemic Heart Disease. *ACS Omega* **3**, 5918-5925 (2018).
58. Zhang, S.*, et al.* CCL20 secreted by KRT15(high) tumor Cells promotes tertiary lymphoid structure formation and enhances anti-PD-1 therapy response in HPV(+)HNSCC. *Cell Death Dis* **17**, 150 (2025).
59. Zhao, Z.*, et al.* Hepatitis B virus promotes its own replication by enhancing RAB5A-mediated dual activation of endosomal and autophagic vesicle pathways. *Emerg Microbes Infect* **12**, 2261556 (2023).

# Methods

**Assembly of the virus–host interaction resource**

We assembled a curated virus–host protein–protein interaction (PPI) resource by integrating records from IntAct, BioGRID, VirHostNet and VirusMentha. To focus on experimentally supported physical associations, we retained only human–virus interaction pairs annotated as physical interactions and excluded ambiguous categories such as colocalization and genetic interaction. Because the evidence systems and annotation schemes differed substantially across repositories, we assigned interaction confidence within each source rather than by applying a single pooled criterion across all databases.

For IntAct, interaction confidence was stratified using the database-provided Molecular Interaction (MI) score, a quantitative estimate of confidence in a given interaction derived from a normalized and weighted count of independent interaction evidence and associated experimental methods. In the IntAct implementation, the score reflects the amount and quality of supporting evidence, including the interaction detection method, interaction type and number of supporting publications. Interactions with $MI > 0.6$ were classified as high confidence, whereas interactions with $0.4 < MI <= 0.6$ were classified as medium confidence. For BioGRID, low-throughput physical interactions were retained as high confidence, whereas high-throughput records were retained only when supported by additional validation and were assigned to the medium-confidence tier. For VirHostNet, binary assays were treated as high-confidence evidence, whereas complex-oriented assays were retained as medium confidence. For VirusMentha, only multi-validated records were retained as high confidence because the available annotation did not support a comparable medium-confidence tier.

After source-specific filtering, redundant records were collapsed with priority assigned to the highest-confidence annotation available for each interaction pair. This procedure yielded a unified virus–host interaction atlas comprising high-confidence and medium-confidence subsets, which were subsequently used for model development, fine-tuning and benchmarking.

**Construction of negative samples**

To avoid overly optimistic performance estimates arising from naïve random sampling, we constructed a conservative hard-negative set using an MMseqs2-based clustering strategy. Human and viral protein sequences were clustered at a sequence similarity threshold of 0.6. For each viral protein, we excluded its known human binding partners and human proteins belonging to the same clusters as those binders, thereby reducing contamination of the negative pool by close homologs of known positives. We then removed candidate negatives whose local sequence identity to any positive sample exceeded 0.6. From the remaining candidate pool, we sampled negatives at a 1:10 positive-to-negative ratio. This design minimizes the inclusion of trivially separable negatives, thereby requiring the model to distinguish experimentally validated binders from non-binders that are sequence-dissimilar yet biologically plausible.

**Dataset partitioning and leakage control**

For the high-confidence dataset, we allocated 10% of interaction pairs to a validation set, 10% to a test set, and used the remaining pairs for model training and fine-tuning. Training and test partitions were separated by enforcing a maximum viral sequence identity of 0.6 between the two sets; that is, no test viral sequence shared more than 60% sequence identity with any viral sequence in the training set. This design reduced viral leakage and enabled evaluation on sequence-dissimilar viral proteins rather than on closely related viral homologs.

Additional leakage-control procedures depended on the benchmark design. In the external coronavirus benchmark, overlapping interaction pairs were removed from the training set before evaluation. Accordingly, this analysis represented a de-leaked interaction-level validation rather than a fully virus-held-out setting, because coronavirus proteins remained well represented in currently available virus–host interaction resources.

**Model architecture**

ViraHinter was implemented as a dual-modal framework that coupled a structure-generation branch with a sequence-representation branch and routed the fused features to both structure and interaction heads. The model took paired human and viral protein sequences as input.

In the structural branch, input sequences were processed using an IntFold backbone initialized from open-source pretrained weights. The structural backbone remained fully frozen during virus–host training. It generated structure-informed single and pair representations for the human–virus pair, which were then used by the downstream interaction module. Consistent with the architecture described in the main text, the structural branch followed a Pairformer-based design and provided intermediate representations for both complex-structure modeling and interaction scoring.

To focus the interaction predictor on intermolecular compatibility rather than monomeric folding alone, we masked intra-chain regions of the pair representation before feature aggregation so that the downstream predictor emphasized inter-chain signals encoding interface compatibility. In parallel, the sequence branch extracted protein language model embeddings using ESM2-150M for the human and viral proteins. We fused these sequence-derived features with the structure-derived representations and refined them through the interaction-prediction module to generate the final interaction probability.

**Input features and MSA generation**

The inputs to ViraHinter comprised paired human and viral protein sequences. We generated multiple-sequence alignments (MSAs) using the same search strategy, databases and database versions as those used for AF3. Specifically, MSA generation followed the ColabFold pipeline with Jackhmmer-based sequence retrieval. By matching the upstream MSA generation procedure to AF3, we reduced the impact of inconsistent alignment construction and maintained a fairer comparison between ViraHinter and AF3-based structural baselines.

**Training procedure**

We initialized the structural branch from the open-source IntFoldV1 weights and trained the model directly on curated virus–host interaction data. Training proceeded in two main stages: adaptation on the broader medium-confidence virus–host atlas, followed by fine-tuning on the high-confidence subset. This design transferred structure-aware representations to the virus–host setting while avoiding reliance on an intermediate generic PPI corpus.

We trained the model on 96 NVIDIA A100 GPUs with a per-GPU batch size of 4. Optimization was performed using AdamW with a learning rate of $1 \times 10^{-4}$ for 30,000 training steps. These settings were used for the main virus–host training runs reported in this study.

**Baseline methods and evaluation protocol**

We benchmarked ViraHinter against AF2 RF2-PPI, AF3 and RF2-Lite using the evaluation settings defined in the corresponding benchmark studies. For fair comparison, rather than applying a single in-house training split to all external evaluations we retrained ViraHinter on the original training data provided by each benchmark. For baseline methods, we used the results reported in the corresponding original publications whenever those results were available. For methods or benchmarks for which published results were not available, we performed inference using publicly available model weights. AF3-based comparisons used the native AF3 confidence score for ranking. The main comparison included the standardized RF2-PPI benchmark under both 1:1000 and 1:10 positive-to-negative ratios, as well as interface-stratified evaluation across weak (≤10 residues), medium (11 to 50 residues) and strong (>50 residues) interaction regimes.

We additionally evaluated ViraHinter on other human-virus benchmarks with distinct split designs, including HVIDB[60], DeepGNHV/C2Virus[61] and an ViraHinter-specific evaluation set. Because the HVIDB split did not fully eliminate viral leakage, we interpreted it as a relatively permissive benchmark. We placed greater emphasis on evaluation settings with stricter separation criteria, including the DeepGNHV/C2Virus benchmark and our ViraHinter-specific split. We used precision–recall analysis and area under the precision–recall curve (AUPR) as the primary evaluation metrics because of the strong class imbalance in proteome-scale interaction screening.

**External coronavirus benchmark and evidence-aware validation**

To assess ranking behavior on an independent coronavirus-focused dataset, we assembled an external benchmark derived from previously reported host–coronavirus interaction resources spanning SARS-CoV-1, SARS-CoV-2 and MERS-CoV. After standardization, we removed records for which the corresponding protein sequences could not be retrieved, as well as pairs containing sequences that exceeded the model's length constraints. The final benchmark contained 958 virus–host protein pairs. To minimize information leakage, any interaction pairs overlapping this benchmark were removed from the training set before evaluation.

Because the coronavirus benchmark was compiled from interactions that were already relatively well supported experimentally, we adopted a stricter training definition for this analysis. Specifically, only the high-confidence subset of the curated IntAct dataset was used as positive training data. Medium-confidence IntAct interactions were not treated as positive training examples in this setting; instead, a subset was incorporated as hard negatives during training. This design reduced the risk that moderately supported interactions would blur the decision boundary, encouraged the model to better distinguish highly supported virus–host interactions from weaker or more ambiguous associations, and ensured a clearer separation between training and evaluation while sharpening the model's sensitivity to high-evidence interactions.

For orthogonal validation, we queried supporting evidence from BioGRID rather than from IntAct, which served as the primary source of positive training interactions in this setting. Benchmark pairs were then stratified into evidence categories, including no qualifying support, high-throughput support and low-throughput support. This evidence-aware framework enabled us to test whether the top-ranked predictions preferentially aligned with independently supported coronavirus–host interactions.

To further explore the interaction landscape of the highest-confidence predictions, we examined the top 100 ranked human gene interaction pairs and focused on host proteins that were initially linked to only one or two viral species. For these candidates, we used ViraHinter to infer interactions with additional viral proteins beyond those supported in the benchmark. Based on the resulting predictions, we identified human proteins whose inferred viral interaction spectrum expanded to include a larger number of viruses and visualized these relationships as virus–host PPI networks. This analysis provided a complementary view of potentially shared host factors and highlighted candidate proteins that may participate in broader cross-viral interaction programs.

**Multi-stage screening and composite reranking for the influenza proteome-wide analysis**

Because the influenza analysis involved proteome-scale screening against the full human protein repertoire, we adopted a multi-stage candidate selection strategy to make large-scale evaluation computationally tractable. We first trained a lightweight sequence-only model on the virus–host interaction dataset. This model used the same sequence-module architecture as the sequence branch of ViraHinter but excluded the structure module and was used solely for rapid prescreening. For each influenza subtype, all candidate virus–host pairs formed between the viral proteins of that sequence and the human protein library were first scored by this sequence-only model, and the top 10,000 candidate pairs were retained. This yielded 30,000 prescreened pairs across the three influenza subtypes.

These prescreened pairs were then evaluated with the full ViraHinter model to obtain joint interaction predictions and corresponding structural outputs. For each subtypes, we selected the top 100 candidate pairs based on the full-model interaction score. To prioritize candidates for downstream structural and functional validation, we further applied an empirical composite reranking score:

Score = (ViraHinter score × 0.4) + (ipTM × 0.4) + (Sequence score × 0.2).

This reranking scheme was designed to integrate three complementary signals. The ViraHinter score captured the interaction confidence assigned by the full structure-aware model, whereas ipTM reflected the internal confidence of the predicted complex structure. The sequence score was included as an orthogonal signal derived from the lightweight prescreening model. Incorporating this term allowed sequence-based support to be retained during final prioritization, which was particularly useful in a proteome-scale setting in which some candidates could be sensitive to uncertainty in structure prediction alone.

Greater weight was assigned to the two structure-aware terms because the final prioritization was intended to favor candidates supported both by strong predicted interactions and by more reliable complex models. The sequence-based score was therefore included as a lower-weight complementary criterion rather than as a dominant determinant of the final ranking. Accordingly, this composite score was used as a practical reranking heuristic for candidate selection and downstream analysis, rather than as a separately trained or statistically optimized predictive model.

**Statistical analysis**

Model performance was evaluated primarily using precision–recall curves and AUPR. For the evidence-aware analyses **(Fig. 3)**, prediction probabilities were stratified by BioGRID evidence class and compared across groups; overall differences in score distributions were assessed using the Kruskal-Wallis test. The relationship between prediction scores and independent evidence support was further quantified using Spearman rank correlation. To evaluate ranking behavior, we performed enrichment analyses on the top-ranked predictions by calculating the proportion of pairs from each evidence group within the top 10% of model predictions and normalizing these values by the corresponding background frequencies in the full benchmark. We additionally quantified enrichment across a range of ranking thresholds by computing enrichment-over-background curves for supported interaction categories, including any qualifying evidence, low-throughput-supported pairs, and pairs supported by both high- and low-throughput evidence. These analyses were used to determine whether independently supported interactions were preferentially concentrated among the highest-scoring model predictions.

We summarized model performance primarily using precision–recall curves and AUPR. For evidence-aware analyses, we compared prediction-score distributions across evidence groups and assessed the relationship between model scores and supporting evidence counts using rank-based correlation analysis. We further performed enrichment analyses across top-ranked prediction bins and ranking thresholds to determine whether independently supported interactions accumulated among the highest-scoring predictions.

**Sequence alignment and identity analysis**

To establish the evolutionary divergence between the influenza A virus (IAV) subtypes evaluated in the ViraHinter framework, we performed pairwise sequence comparisons of the 10 core viral proteins (PB2, PB1, PA, HA, NP, NA, M1, M2, NS1, and NEP) across the seasonal A/Nagasaki/07N020/2008 (H1N1) strain, the seasonal A/China/JN22022/2022 (H3N2) strain and an avian-origin A/chicken/Korea/KNUGJ09/2009 (H9N2) strain. The complete set of viral protein sequences encoded by H1N1, H3N2, and H9N2, as used for the large-scale interactome screening, is provided in **Extended Data Table 7.** Pairwise sequence identity was assessed using the Basic Local Alignment Search Tool (BLAST). Specifically, BLASTN was employed to evaluate nucleotide-level similarity among viral segments for genomic lineage confirmation. The amino acid sequence identities **(Fig. 4a)** reflect evolutionary constraints at interaction interfaces, and the resulting identity matrix was used to aassociate evolutionary conservation with predicted protein–protein interaction (PPI) similarity across virus subtypes.

**Software and reproducibility**

We implemented preprocessing, training and downstream analyses using standard Python-based scientific computing workflows. MSA generation was performed using the same search procedure and database setup as AF3.

## Data availability

The studies referenced in our case analyses are available from previously published work[5]. Associated datasets are provided in these publications. The amino acid sequences of the H3N2 strain used in this study are deposited in the National Microbiology Data Center (https://nmdc.cn; accession number: NMDCP0018772-NMDCP0018783). Data generated and processed in this study are publicly available at the following repository: Google Drive folder.

## Code availability

The code for this study will be made publicly available on GitHub soon.

## References

60. Yang, X., *et al.* HVIDB: a comprehensive database for human-virus protein-protein interactions. *Brief Bioinform* **22**, 832-844 (2021).
61. Jiang, L., *et al.* Graph neural network integrated with pretrained protein language model for predicting human-virus protein-protein interactions. *Brief Bioinform* **26**(2025).

## Acknowledgements

This work was funded by the Prevention and Control of Emerging and Major Infectious Diseases-National Science and Technology Major Project (2025ZD01903701).

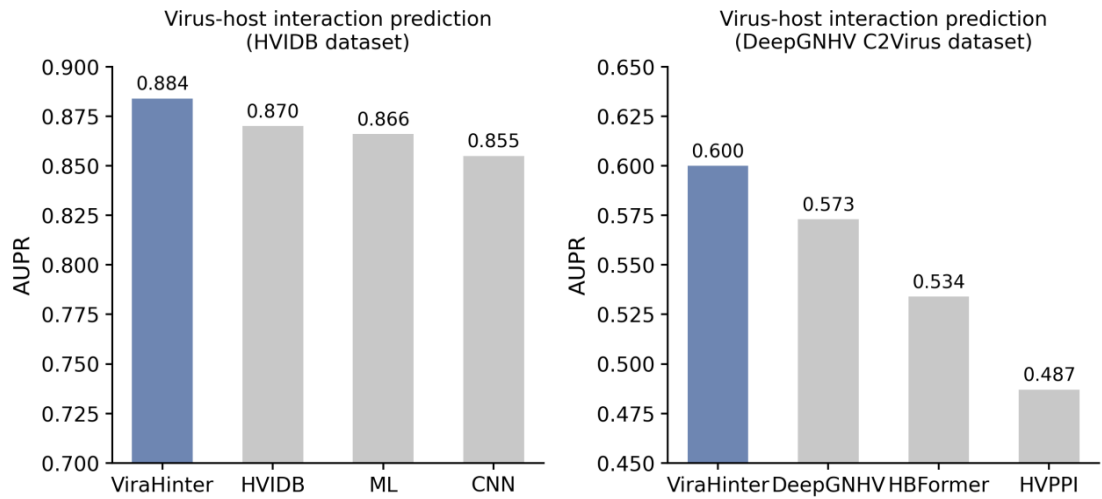

**Extended Data Fig. 1 | Performance of ViraHinter on additional virus-host interaction benchmarks.**
**a,** Performance comparison on the HVIDB dataset. ViraHinter achieved the highest area under the precision–recall curve (AUPR, 0.884), outperforming the HVIDB baseline, a conventional machine-learning model and a CNN-based model. **b,** Performance comparison on the DeepGNHV/C2Virus dataset. Under this stricter benchmark, ViraHinter again achieved the best performance (AUPR, 0.600), exceeding DeepGNHV, HBFormer and HVPPI. Together, these results show that ViraHinter maintains strong predictive performance across independent human–virus interaction benchmarks with different partitioning schemes and levels of viral leakage control.

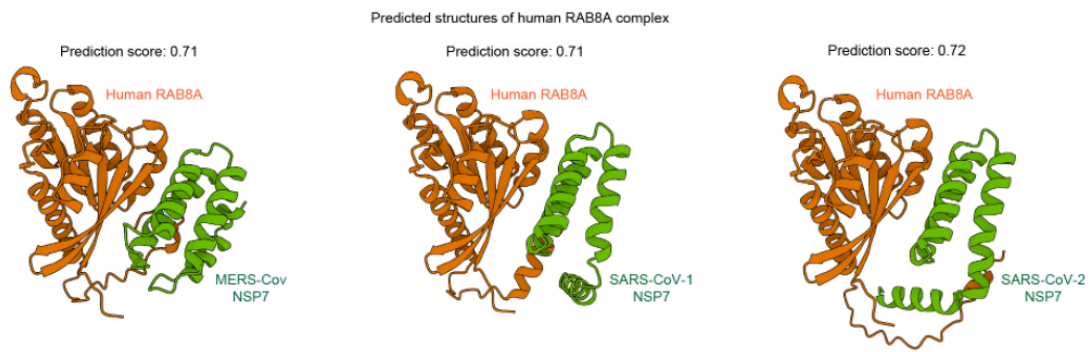

**Extended Data Fig. 2 | Conserved predicted binding mode of RAB8A with NSP7 from three pathogenic coronaviruses.**
Predicted complex structures of human RAB8A with NSP7 from MERS-CoV, SARS-CoV-1 and SARS-CoV-2. Human RAB8A is shown in orange and viral NSP7 proteins are shown in green. Prediction scores are indicated above each model. Despite sequence divergence among NSP7 orthologues, the three complexes adopt highly similar docking geometries, with the viral proteins converging on a common surface region of RAB8A. These structural models support the notion that the NSP7–RAB8A interaction represents a conserved coronavirus–host interface and provide a mechanistic basis for the shared host dependency identified by ViraHinter.

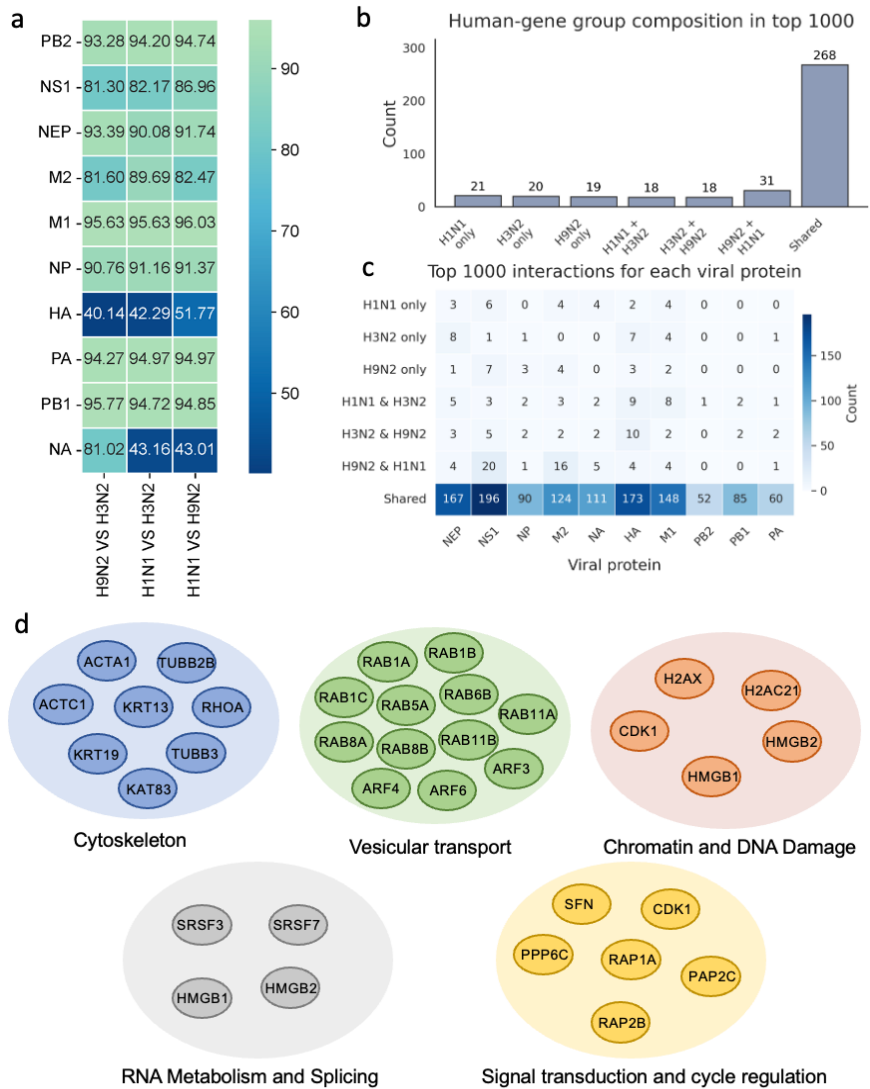

**Extended Data Fig. 3 | Sequence divergence, shared host targeting and functional organization of the predicted influenza A virus interactome.**
**a,** Pairwise amino-acid sequence identity of the 10 core viral proteins across the three influenza A virus strains used in this study (H1N1, H3N2 and H9N2). Values indicate the sequence identity for each viral protein between strain pairs, illustrating substantial divergence for several proteins, particularly HA, M2 and NA. **b,** Composition of host-gene groups in the top 100 ranked candidates. Predicted host factors are partitioned into strain-specific groups, pairwise-shared groups and a three-strain shared group, showing that a large fraction of host genes are targeted by all three strains. **c,** Distribution of the top 100 predicted interactions across individual viral proteins, stratified by host-gene sharing category. Shared host factors are enriched among multiple viral proteins, whereas subtype-specific interactions are more restricted, revealing both conserved and protein-specific targeting patterns. **d,** Functional categorization of the 33 host factors shared by all three influenza strains. These common host targets cluster into five major biological modules, including cytoskeleton, vesicular transport, chromatin and DNA damage, RNA metabolism and splicing, and signal transduction and cell-cycle regulation, highlighting the core cellular systems recurrently targeted across divergent influenza viruses.